\documentclass[10pt,twocolumn, twoside]{IEEEtran}
\usepackage{tabularx}
\usepackage{graphicx}
\usepackage{subfigure}
\usepackage{mathrsfs,amsmath}
\usepackage{color}
\usepackage[utf8]{inputenc}
\usepackage{amsmath,amssymb,epsfig,theorem,url,cite,bm,soul}
\usepackage{multirow}
\usepackage{rotating}
\usepackage{wrapfig }
\usepackage{times}
\usepackage{fancyhdr,graphicx,amsmath,amssymb,enumitem}
\usepackage[ruled,vlined]{algorithm2e}
\include{pythonlisting}

\usepackage{soul}
\usepackage{color}
\usepackage[usenames,dvipsnames]{xcolor}
\usepackage{tablefootnote}

\usepackage{amssymb}\usepackage{pifont}

\begin{document}

\title{Video Source Characterization Using Encoding and Encapsulation Characteristics}

\author{\IEEEauthorblockN{
Enes Altinisik,
H\"usrev Taha Sencar,
Diram Tabaa
}
}

\maketitle   

\begin{abstract}

\textcolor{black}{
We introduce the use of video coding settings for source identification 
and propose a new approach that incorporates encoding and encapsulation aspects of a video.} To this end, a joint representation of the overall file metadata is developed and used in conjunction with a two-level hierarchical classification method.
At the first level, our method groups videos into metaclasses considering several abstractions that represent high-level structural properties of file metadata. 
This is followed by a more nuanced classification of classes that comprise each metaclass.
The method is evaluated on more than 20K videos obtained by combining four public video datasets.
Tests show that a balanced accuracy of 91\% is achieved in correctly identifying the class of a video among 119 video classes.
This corresponds to an improvement of 6.5\% over the conventional approach based on video file encapsulation characteristics. 
\textcolor{black}{Analysis performed on a large, unlabeled video set also confirmed the aptness of our approach.} \textcolor{black}{To further demonstrate the versatility of encoding parameters, we consider attribution of partial video files where file metadata is not available. 
Our results show that, even in this limited setting that is intrinsic to forensic file recovery, an identification accuracy of 57\% can be achieved through the use of a subset of encoding parameters estimated from coded video data.}

\end{abstract}

\section{Introduction}

An essential requisite for media forensics is the ability to track the provenance of multimedia content.
Such a capability may answer several questions about the origin of an image or video with differing levels of specificity.
At the one end, the focus may be on identifying whether a given media is camera-captured or a deepfake, i.e., synthesized by a deep neural network, 
\cite{tolosana2020deepfakes, verdoliva2020media}.
At the other end, it may be about attributing the media to the particular device that generated it \cite{chen2008determining,dirik2008digital}.
In between, the inquiry may concern some class-level characteristics of the media at hand such as the brand and model of the acquiring camera 
\cite{ho2015forensic,junior2019depth},
the social media platform it is downloaded from \cite{amerini2017tracing, amerini2017dealing}, or the deep learning algorithm used for generation \cite{marra2019gans, yu2019attributing}. 

Forensic methods proposed to address different instances of the attribution problem mainly adopt a content-based analysis approach to uncover relevant forensic traces. 
Among these, source camera attribution based on a sensor's photo-response non-uniformity (PRNU) is now very well established.
The most important factor underlying the effectiveness of this method is the multi-dimensional nature of the PRNU,
which depends on the number of photosensitive elements (i.e., pixels) in a sensor, thereby yielding a highly discriminative signal.
Considerable success is also reported in source attribution settings that involve a classification problem with fewer classes such as detecting deepfake media and  
provenance tracing in online social networks.

In contrast to these successes, reliably identifying the camera model that produced a media remains a challenge. 
Essentially, camera-model attribution requires the extraction of distinctive characteristics that can distinguish a large number of source classes. 
In this regard, characterizing specifics of individual processing steps applied in-camera and during post-processing (such as color filter array configuration \cite{kirchner2010efficient,choi2011cfa}, demosaicing algorithm \cite{bayram2006improvements,swaminathan2007nonintrusive,chen2015camera}, lens-related distortions \cite{san2006automatic,van2007identifying,yu2011toward}) or the better performing data-driven modeling methods \cite{bondi2016first, marra2017study} do not yield sufficient discriminative information to distinguish between many camera models. 
Improving the accuracy of this task crucially depends on identifying and combining several such model-dependent characteristics.

In source identification, an alternative approach to content-based analysis exploits the representation aspect of image and video data. 
Multimedia data is almost always stored and transferred in some compressed file format, therefore it is accompanied by rich metadata and information that govern its organization to ensure successful reconstruction.
Moreover, due to variations in implementation of formatting standards and the choice in parameters, 
attributes of a video file are expected to vary across camera models, as well as between video processing software and firmware,  
while remaining largely invariant among media generated through the same process.
Combined with the fact that such information is readily available, without the need for a complicated extraction procedure, file structure and the associated metadata potentially offer a high discrimination capability.

Videos typically include a large amount of audio-visual data with strict rendering requirements, and
video coding is a quite sophisticated process involving many parameters.
Therefore, components and structure of a video file provide a suitable basis for source identification.  
This was initially observed by Gloe et al. \cite{gloe2014forensic} who examined videos in several file formats generated by many cameras and editing tools. 
Based on this observation, the subsequent work focused on file container formats and exploited the fact that file data is packed in a sequence of compartments, known as {\em boxes} in MPEG 4-based file formats and {\em resources} in AVI format, to identify camera model/brand and detect video tampering.

In \cite{song2016integrity}, Song et al. proposed using the sequence of AVI metadata fields in a video as a signature of a camera model.
The following research mainly focused on developing a representation for video file metadata.
For this, Iuliani et al. \cite{iuliani2018video} proposed using an ordered listing of MP4 file metadata fields and their values as a representative feature vector.
Yang et al. \cite{yang2020efficient} modified this initial representation by disregarding order information and decoupling fields from their values. 
To improve the robustness of the representation against insertion and deletion of fields, Gelbing et al. \cite{gelbing2021video} introduced an alternative enumeration of fields. Based on a similar representation, Xiang et al. \cite{xiang2021forensic} proposed another improvement in how non-categorical values are expressed and performed linear discriminant analysis to further reduce the dimensionality of the feature vector.

\textcolor{black}{
The above studies have so far focused on exploiting the rich metadata describing the structure of the container file format. 
This line of work, however, overlooked an equally important aspect of video file generation: the encoding and decoding setting of a video.
Video coding is an indispensable component of camera imaging pipelines as well as video editing tools due to  
impractically high storage and bandwidth requirements of handling raw video data.
}
\textcolor{black}{
In fact, encoding operation provides a large freedom in the choice of parameter values needed for the operation of a video coder-decoder (codec).
In practice, however, fine tuning parameters individually for each video requires additional computation (at the encoder side) while yielding a dispensable gain in compression ratio.
Therefore, manufacturers and developers have a tendency to use predetermined parameter settings as evidenced in the case of JPEG image encoders deployed by cameras and photo editing software \cite{uzun2015carving,kee2011digital}.
}
\textcolor{black}{
In addition, the encoder used for compression and the file format governing the arrangement of coded data can be selected independently.
Hence, the use of codec parameters provides a complementary, and not competing, capability to existing approaches. }

\textcolor{black}{
According to the Video Developer Survey 2019 \cite{bitmovin}, 91\% of the developers use H.264 video coding standard.
In concert with this, analysis of more than 100K user-uploaded videos revealed that 99.6\% of those were encoded using H.264 format \cite{altinisik2021automatic}.
Thus, H.264 coding characteristics may serve as an important basis for video source identification. 
Motivated by this, in this work, we introduce the use of H.264 video sequence headers that contain sets of codec parameters for video source attribution. 
}

\textcolor{black}{
\textbf{Contributions.} The main contributions of this paper are as follows.}
\begin{itemize}[leftmargin=*]
  \item \textcolor{black}{We determined the parameter sets used by H.264 codecs of 109 camera models, three video editing tools, and four social media platforms
  by examining videos contained in four public datasets, including the VISION \cite{shullani2017vision}, ACID \cite{acid}, SOCRatES \cite{galdi2019socrates} and EVA-7K \cite{yang2020efficient} datasets.
  The variation of parameters within and across these video source classes is evaluated. } 

  \item \textcolor{black}{We compare and contrast the distinctiveness of H.264 coding and MP4 encapsulation characteristics.
  To also demonstrate their complementary nature, we present a joint representation by combining these characteristics in a file-metadata tree.
  We then introduce a hierarchical method that first exploits the topological properties of the resulting tree representation followed by the use of field and parameter values to provide a more scalable approach to source attribution.}
      
  \item \textcolor{black}{The method is validated on the largest dataset used in any study of this nature, comprising 20,153 videos of 119 source classes,   The experiment results demonstrate that a balanced accuracy of 91\% is achieved in distinguishing these video source classes.
  Comparisons performed by implementing three state-of-the-art methods utilizing MP4 file metadata, \cite{yang2020efficient, gelbing2021video, xiang2021forensic}, show that incorporation of encoding parameters as part of file metadata yields an overall improvement of 6.5\% in accuracy.} 

 \item \textcolor{black}{We identified a new MP4 file metadata parser \cite{GPACNigh31} as an alternative to the commonly used parser by previous work \cite{iuliani2018video,yang2020efficient,gelbing2021video}.
 Our tests showed that associated features obtained using this parser are more comprehensive and allow better discrimination of a video's source class. }

  \item\textcolor{black}{To better assess the scalability of our approach, we extend analysis to an open setting by examining file-metadata tree representations of 92K videos obtained from almost 15K 
  user accounts in the lbry.com video sharing platform. 
  The diversity in the encoding and encapsulation behavior is evaluated in comparison to that seen in the combined public datasets.
    }

  \item \textcolor{black}{We extend the use of codec parameters to attribute incomplete video files.
  Partial files are encountered frequently when performing forensic data recovery from storage media during which file metadata may be missing or unlocatable. 
    In such cases, some of the encoding parameters can be determined through analysis of coded frame data as introduced in \cite{altinisik2021automatic}.
  We demonstrate that by using those estimated parameters a balanced accuracy of 57\% can be achieved in attributing a video to its source classes.}

\end{itemize}

The rest of the paper is organized as follows:
In the next section, we describe the video generation process from the perspective of H.264 encoding and MP4 file encapsulation due to their prevalence.
\textcolor{black}{Section \ref{sec:graph} provides an overview of methods used for representing file metadata.}
The distinctiveness of coding parameters is analyzed in Sec. \ref{sec:Distinctiveness}.
Our representation for file metadata that incorporates both aspects of a video file and the details of the two-level hierarchical classification method are introduced in Sec. \ref{sec:method}.
Identification accuracy results obtained on the combined dataset under several test settings are provided
along with our observations of videos acquired from lbry.com in Sec. \ref{sec:results}.
Finally, we discuss our findings in Sec. \ref{sec:discussion} and present our conclusions in Sec. \ref{sec:conclusion}.

\section{Video File Generation}
A video file is generated in a camera at the last processing stage of the acquisition pipeline.
At this stage, the main objective is to reduce the size of frames created by the earlier processing stages and package the coded data before it can be saved. 
A video encoder performs one part of this task by removing the temporal and spatial redundancy in and between successive frames.
Several video coding standards with improved coding performance have been introduced over time, such as MPEG-1, \textcolor{black}{MPEG-2 Part 2} (H.262),
\textcolor{black}{MPEG-4 Part 2 (compatible with H.263)}, \textcolor{black}{MPEG-4 Part 10 (AVC or H.264)}, and \textcolor{black}{MPEG-H Part 2 (HEVC or H.265)}. 
In addition, several alternative encoding formats that are inspired by H.264 and H.265 standards have been developed independently, including VP8/9, VC, and AV1.

The coded video frames are then combined with other essential data, including coded audio segments, encoding parameters, subtitles, etc., in a container file.
File container formats are essentially optimized for different use cases, i.e., streaming or playback, and support a range of audio and video codecs.
For playback videos, the ISO/IEC 14496-12 (MPEG-4 Part 12) standard specifies a general media file format \cite{ISOISOIE63}, and many widely used file formats, such as MP4, 3GP, and MOV, are based on it. 
Besides, there are several open and proprietary container formats such as MKV and AVI.
For video streaming, the MPEG Transport Stream (TS) standard specifies a container format for encapsulating packetized data.
This forms the basis of the most commonly used streaming formats such as HTTP Live Streaming (HLS) and MPEG-DASH \cite{bitmovin}.

Among the existing file encoding formats, H.264 is by far the most widely used codec today \cite{bitmovin, altinisik2021automatic}.
In terms of file container formats, despite a great diversity of options, mobile phones predominantly use file formats that extend over the MPEG-4 standard\cite{huaman2020authentication}.
Since smartphones serve as the de facto camera, methods proposed for the characterization of video files mainly focused on MP4 files.
Therefore, we examine the overall structure and content of H.264 encoded videos contained in MP4-like file formats, \textcolor{black}{which include MP4 (de-facto standard for Android devices) and MOV (used by Apple devices)},  more closely.

\subsubsection{H.264 Coding Parameters}
H.264 encoding follows a block-based coding approach where compression of picture blocks involves the key steps of prediction, (in-loop) deblocking filtering, transformation, quantization, and entropy coding \cite{H264}.
These processing steps are governed by a set of parameters that are dynamically determined by the encoder to attain a target compression bitrate and picture quality. 
The same parameters are required at the decoder to recreate the video.
Therefore, they are stored along with the coded data. 
The H.264 video coding parameters can be grouped into three sets:

\textit{Sequence Parameter Set (SPS):}
The SPS includes 38 parameters and applies to a sequence of pictures that are encoded in an interdependent manner. 
Some of these parameters specify general characteristics of the video, such as resolution, bitrate, and frame rate, so that the decoder can ensure it has sufficient resources to handle the video bitstream. 
Another subset includes parameters that define the bit widths of several variables needed for parsing the bitstream. 
A third subset of parameters provides values for variables such as the width and height of video pictures and the number of reference frames that must be kept in memory during decoding.  
The remaining subsets serve as sequence metadata and contain information like SPS ID (as multiple SPSs can be present) and a bit flag to denote whether {\em video usability information} parameters are included as part of the header.

\textit{Picture Parameter Set (PPS):}
A PPS contains 25 parameters that specify the entropy coding method, motion prediction mode, baseline quantization values, and deblocking filter settings.
Every picture in a sequence may be accompanied with a separate PPS header. 
The PPS, when combined with the SPS, provides all the information needed to reconstruct a picture in a standalone manner.

\textit{Video Usability Information (VUI):}
The VUI parameters are optional and are used in the post-decoding stage to prepare the resulting video sequence for output and display. 
This may include up to 32 parameters and may provide supplemental information about the aspect ratio of coded pictures, the original format of the source video and its color space,  picture output timing to ensure correct play-out speed of the decoded video sequence, and potential bitstream restrictions that help the decoder pick the right computational configuration for decoding.  
When they are present, VUI parameters appear as part of the SPS.

\subsubsection{Structure of MP4-like Files}
In MP4 and other similar container file formats based on the ISO/IEC 14496 Part 12 standard, the basic data structure is referred to as a {\em box} or an {\em atom} following the earlier-defined QuickTime file format.
Accordingly, each box is identified by a four byte ASCII code showing its type which is preceded by a four-byte value indicating the size of the box. 
The boxes in a file are stored in a hierarchical order allowing each box to contain further sub-boxes as well as data values organized in attribute fields. 
Hence the structure of data contained within MP4-like files can be represented as a labeled-tree where internal nodes include boxes labeled by their type and leaf nodes labeled by field-value attributes.
Many different types of boxes appear at the top of the hierarchy, most notably the \texttt{ftype, moov, mdat} boxes.
Among these, the \texttt{moov} box is the most important and has several children nodes containing information about the movie header, audio and video tracks, and references to raw data.  
In terms of the size of contained data, the \texttt{mdat} box takes most of the file size as it contains the coded media data.

\subsubsection{Encapsulation of Coding Parameters} In MP4-like file containers, the \texttt{accv} box nested under the \texttt{moov} box stores a variety of information on the coded bitstream.
Among these, the \textit{AVC Decoder Config Record} contains the needed SPS and PPS headers.
The headers are stored as bit strings which the decoder parses into parameters to initialize itself prior to decoding the video bit stream.  
Our examination revealed that the commonly used parser in extracting MP4 file metadata \cite{sanniesm0} by several researchers (e.g., \cite{iuliani2018video,yang2020efficient,gelbing2021video}) in fact does not extract this information.
Regardless of a parser's behavior, however, expressing encoding parameters in a combined manner as field-attribute values suffers from two main limitations. 
First, this is a sub-optimal way of combining otherwise orthogonal sets of features that relate to coding and packaging aspects. 
Therefore, videos encoded using H.264 but encapsulated into other formats, such as transport streams, cannot be characterized.
Second, and more critically, storing the whole SPS and PPS as a bit string essentially corresponds to combining up to 95 parameters, 
all of which may independently serve as features, into a single feature.
As a consequence of this, a change in any of the parameter values
will result in a completely new value in the \texttt{accv} box field.
Therefore, the difference between camera models cannot be accurately distinguished.

\section{\textcolor{black}{Representations for File Metadata}}
\label{sec:graph}
In the MP4 file format, information is organized in a hierarchy of boxes that either contain other boxes or data organized in attribute fields.
Since boxes may appear in any order, have different boxes nested under them, and vary in their values, their organization and content serve as a means for characterizing the source device that encapsulated a media file.
Several studies have proposed ways of interpreting this tree-structured metadata to obtain a feature representation.
In common, this is realized by extracting all paths of a metadata tree starting at the root node and ending at leaves \cite{iuliani2018video,lopez2020digital,yang2020efficient,gelbing2021video,xiang2021forensic}.
By combining labels of all nodes (i.e., box types) along a path with the field-value pairs at the leaf nodes, a list of path entries is obtained.
As an example, the root-to-leaf path \texttt{moov/mvhd/@duration/17654} indicates that the file contains a nested \texttt{mvhd} box in the parent \texttt{moov} box with a \texttt{@duration} field of value \texttt{17654}.
The observed root-to-leaf paths associated with different video files are then used for discriminating camera models.

\begin{figure}[htbp]
\centering
    \centering
    \includegraphics[width=0.75\columnwidth]{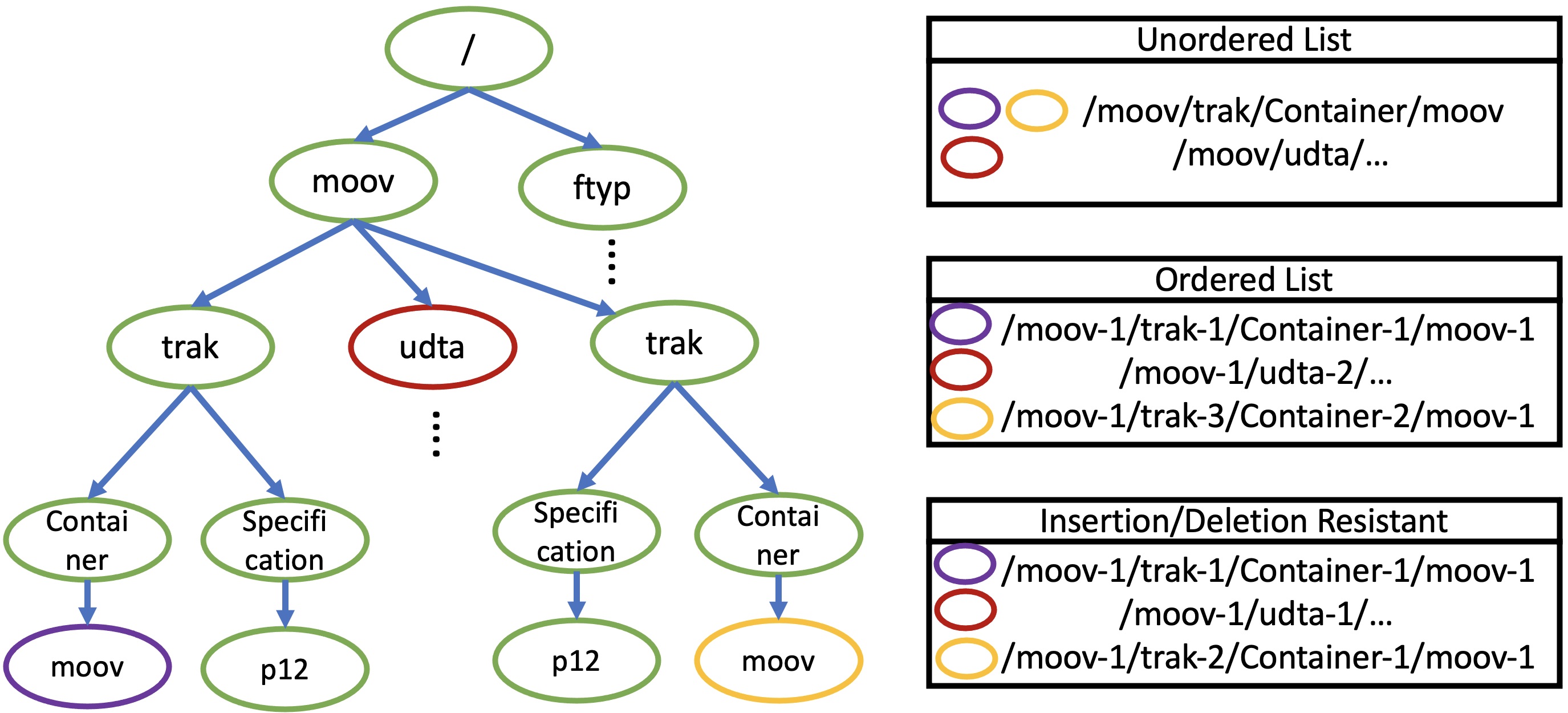}
  \caption{Feature representations corresponding to three mappings of encapsulation metadata to root-to-leaf paths.} 
  \label{fig:represent}
\end{figure}

Several approaches have been proposed to represent file metadata. 
In \cite{yang2020efficient}, extracted root-to-leaf paths are kept as an unordered list, and a sparse feature vector is generated.
The dimensionality of this vector is the same as the number of distinct paths observed over all the dataset where each dimension serves as a binary feature indicating whether or not a particular path is observed. 
This basic representation, however, cannot capture a change in the order of boxes that appear at the top of the tree, 
which will preserve the functionality but imply a change in the encapsulation behavior. 
To address this issue, labels of child nodes of each parent node are further annotated with their order of appearance in the file \cite{iuliani2018video}.
This modification also allowed the easy detection of box deletions and insertions, as all boxes following the inserted or deleted box would attain new labels and yield distinct paths. 
However, a sequential change in labels following a deleted or inserted box also causes otherwise similar paths to be annotated completely differently. 
To better contain such changes in labels, \cite{gelbing2021video} proposed adding order information to only boxes of the same type, thereby an insertion or deletion of a certain type of a box will not affect labels of other types of nodes descending from the same parent.
Figure \ref{fig:represent} shows an illustration of different labeling schemes used in the generation of root-to-leaf paths.

A downside of the above representation is that it yields very high-dimensional, sparse vectors as each distinct path receives its own dimension. 
Xiang et al. \cite{xiang2021forensic} proposed an alternative representation to lower the dimensionality of the feature vector.
For this, root-to-leaf paths are incorporated with paths extending to each node in the tree, i.e., root-to-node paths, while encoding order information only in the labels of the media track (\texttt{trak}) boxes.
Unlike previous approaches that map each distinct path to a separate feature, this approach uses the number of occurrences for each distinct path entry as a feature. 
In addition, root-to-leaf paths are discriminated depending on the type of box field, i.e., whether they take categorical or non-categorical values
with the latter ones assigned their value as features, such as width, height, bitrate, rather than their number of occurrences.
To further reduce the dimension of the resulting feature vector, correlated features are eliminated through a feature selection procedure, and linear discriminant analysis (LDA) is used to ultimately obtain a two-dimensional feature vector.

An important consideration here is the parsing of metadata from the video files.
The flexibility of the MP4 file format allowed manufacturers to accommodate changes in the technology and to address new use cases. 
However, such manufacturer-specific additions also introduce a complication that prevents the design of a universal parser.
In this regard, the MP4 Parser library \cite{sanniesm0} has been a commonly used tool for this purpose \cite{iuliani2018video,yang2020efficient,gelbing2021video}.
It is determined in \cite{xiang2021forensic} that this tool fails to correctly extract the information in the \texttt{ilst} (metadata item list) box due to its non-standard usage.  
Similarly, it is observed that video editing tools incorporate another layer of metadata that requires further interpretation. 
Building a comprehensive parser is not trivial as it requires knowledge of manufacturer-level variations in the metadata. 
Therefore, in this study, we use two publicly available parsing libraries \cite{sanniesm0,GPACNigh31}.

\section{\textcolor{black}{Distinctiveness of Encoding Parameters}}
\label{sec:Distinctiveness}
We explore the distinctiveness of encoding parameters used by different camera models during video coding.  
To this objective, we examined 19,472 H.264 coded videos in the ACID \cite{acid}, VISION \cite{shullani2017vision}, SOCRatES \cite{galdi2019socrates}, and EVA-7K \cite{yang2020efficient} datasets.
These videos are divided into 116 classes that include 109 camera models, three video editing tools, and four social media platforms.  
In all these videos, we identified a total of 535 distinct encoding settings involving 124 unique SPS, 53 unique PPS, and 407 unique VUI headers.
In total, only 18 video classes are observed to not include the optional VUI parameters.
By comparing values of parameters extracted from these videos, we evaluate the degree of distinctiveness exhibited at the class level.

\subsection{Within-Video Variation}
The H.264 standard allows an encoder to vary its parameters during coding of a video sequence. 
Therefore, we first examine how variant SPS and PPS are over the duration of a video. 
Our analysis revealed that except for one class of videos, all videos are encoded using fixed sets of parameters.
That is, a video stream is encoded using the same SPS and PPS specific to that class. This can be attributed to the fact that video coding must be performed in real-time with limited computational resources without much room for optimization. 
For the Sony Cyber Shot DSC-WX350 camera, however, each video is determined to use four different PPSs in their coding.
This essentially indicates that in practice a video file can be characterized by the SPS and PPS used in its generation.

\subsection{Within-Class Variation}
\label{sec:model_var}

Next, we examine the variation of encoding parameters within and across video classes. 
To this objective, we create a dictionary by collating all headers associated with each video class.
This allows us to determine how many distinct headers are used by each video class and reveal the overlap of headers among video classes. 
Such a dictionary is created separately for SPS, PPS, and VUI parameter sets, and considering the combination of the three that defines the overall encoding setting of a video. 

\begin{figure}[htbp]
\centering
    \centering
    \includegraphics[width=0.60\columnwidth]{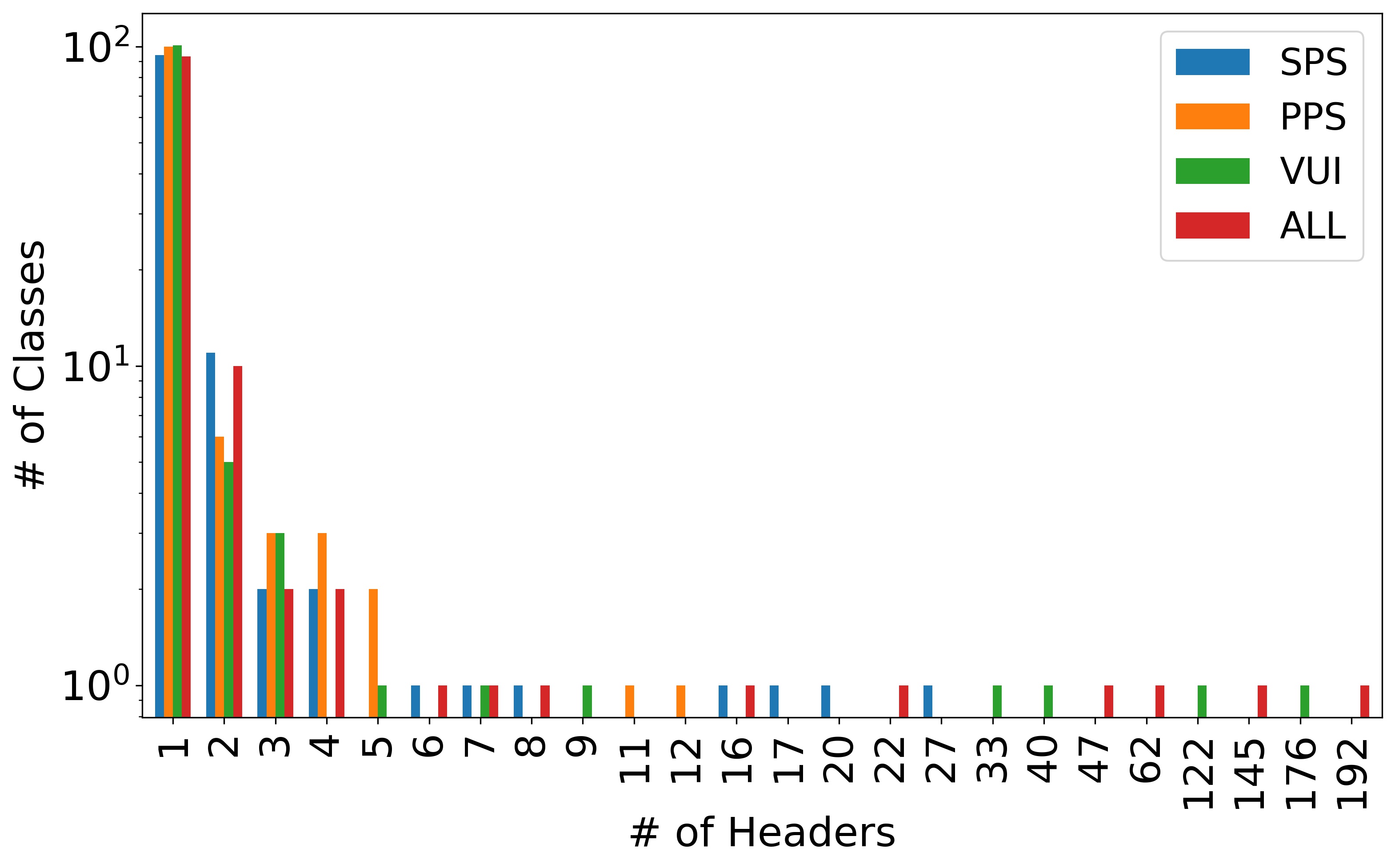}
  \caption{
    Within-class variability measured as the number of classes whose videos are generated using a varying number of distinct headers.
  }
  \label{fig:header2model}
\end{figure}

Figure \ref{fig:header2model} displays a measure of within-class variability in terms of the count of video classes with a specified number of distinct headers used in their generation.
Accordingly, 93 video classes in our dataset are generated using fixed SPS, VUI, and PPS headers.
This finding does not imply that encoding settings are unique to those classes but rather it shows that these video classes don't exhibit any within-class variation. 
All the classes that involve more than eight encoding settings are associated with either editing tools or social media platforms.
Among these, videos sourced from YouTube and Facebook comprise the most diverse classes. 
This is most likely due to these platforms trying to preserve the quality of the original videos by using a compatible encoding setting when re-encoding them.
As for the two video editing tools, Avidemux and FFMPEG, they were configured to carry over the original codec parameters of the source video 
to the edited version.

It is also determined that the VUI parameter set shows higher within-class variation with several classes having more than 30 VUI parameter sets.
Since VUI parameters appear as part of the SPS, this finding implies that treating the SPS string found in the \texttt{accv} box as a feature value by itself results in the creation of many features for the same video class.
This emphasizes the need for decoupling encoding parameters to better differentiate video classes.

\begin{figure}[htbp]
\centering
    \centering
    \includegraphics[width=0.60\columnwidth]{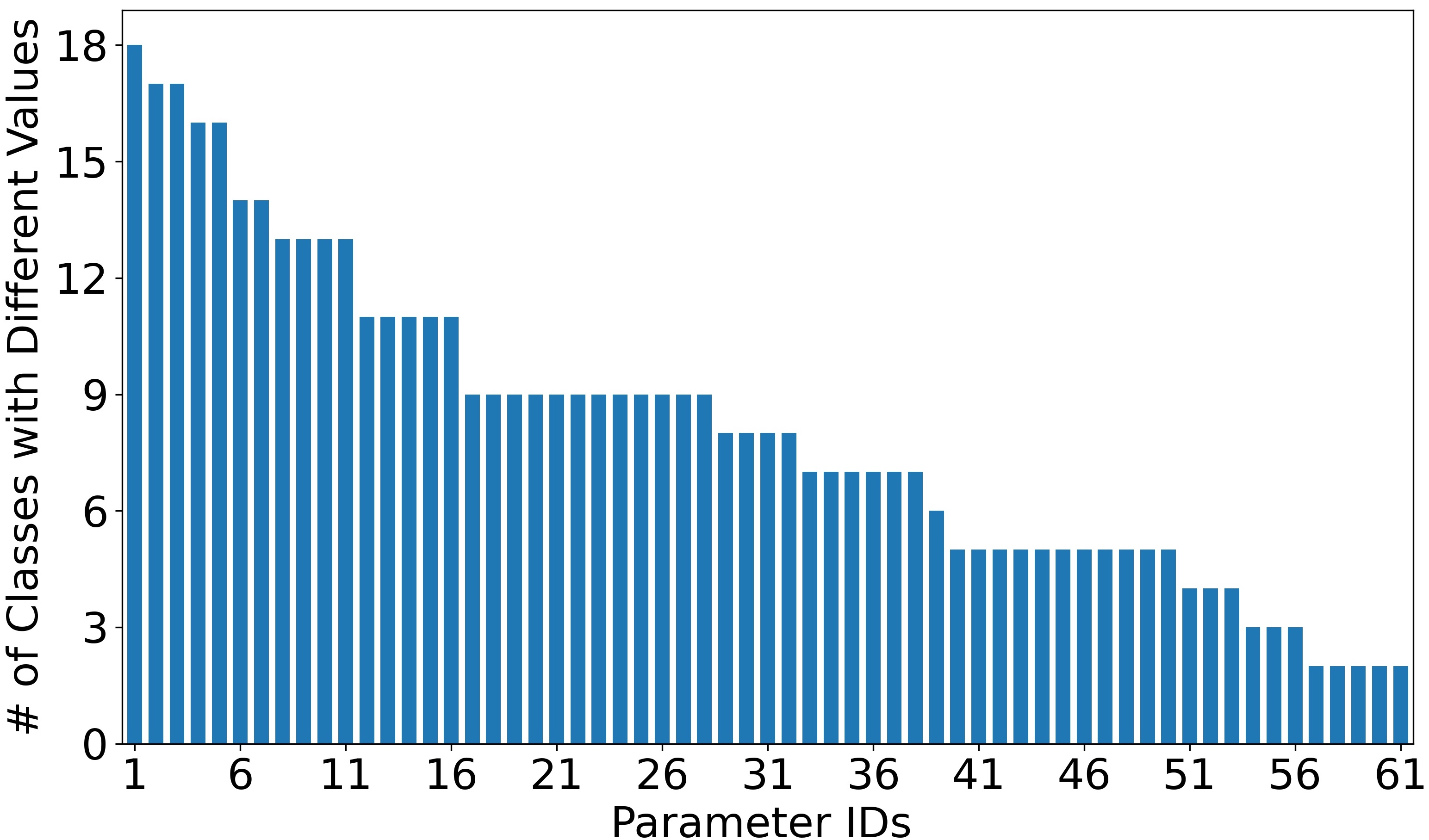}
  \caption{The number of video classes that each encoding parameter is observed to exhibit within-class variation. \textcolor{black}{The mapping between those 61 coding parameters and their IDs is provided in the Appendix.} The remaining 34 parameters (out of 95) only showed between-class variation while remaining same within each video class. }
     \label{fig:parameterLevel}
\end{figure}

When examined at the parameter level, the within-class variation is mostly due to a small number of parameters. 
Out of the 95 parameters involved in video coding, 34 parameters exhibit no variation in any of 119 video classes, i.e., for each video class these parameters take class-specific but fixed values. 
Figure \ref{fig:parameterLevel} shows the number of classes where each of the remaining 61 parameters exhibited some variation.
Accordingly, the parameter that varies most across all video classes is the \texttt{level\_idc} parameter which indicates the quality 
of a video and shows within-class variation in 18 video classes. 
This is followed by four frame resolution-related parameters that take different values in 16 to 17 video classes. 
Overall, we determine that 79 parameters (those excluding the first 16 in Fig. \ref{fig:parameterLevel}) show variation only in 10 or fewer classes. Since parameters that exhibit high within-class variation cannot be intrinsic to a video class, these 79 parameters will be more distinctive in identifying video classes.

\subsection{Between-Class Variation}

\begin{figure}[htbp]
  \begin{minipage}[b]{0.455\columnwidth}
    \centering
    \includegraphics[width=0.98\columnwidth]{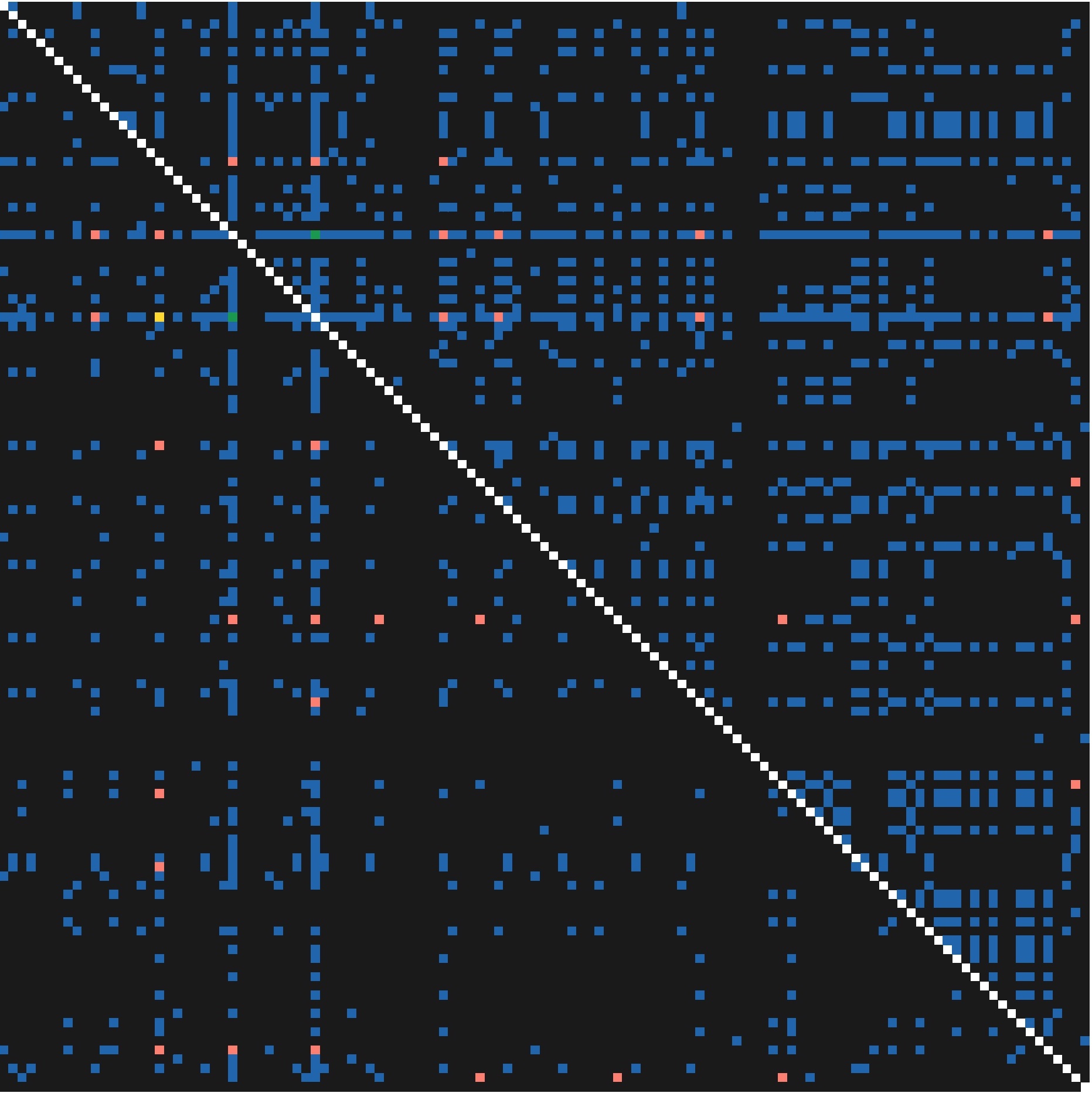}
    \centerline{\scriptsize{(a)}}
  \end{minipage}
  \begin{minipage}[b]{0.545\columnwidth}
    \centering
    \includegraphics[width=0.98\columnwidth]{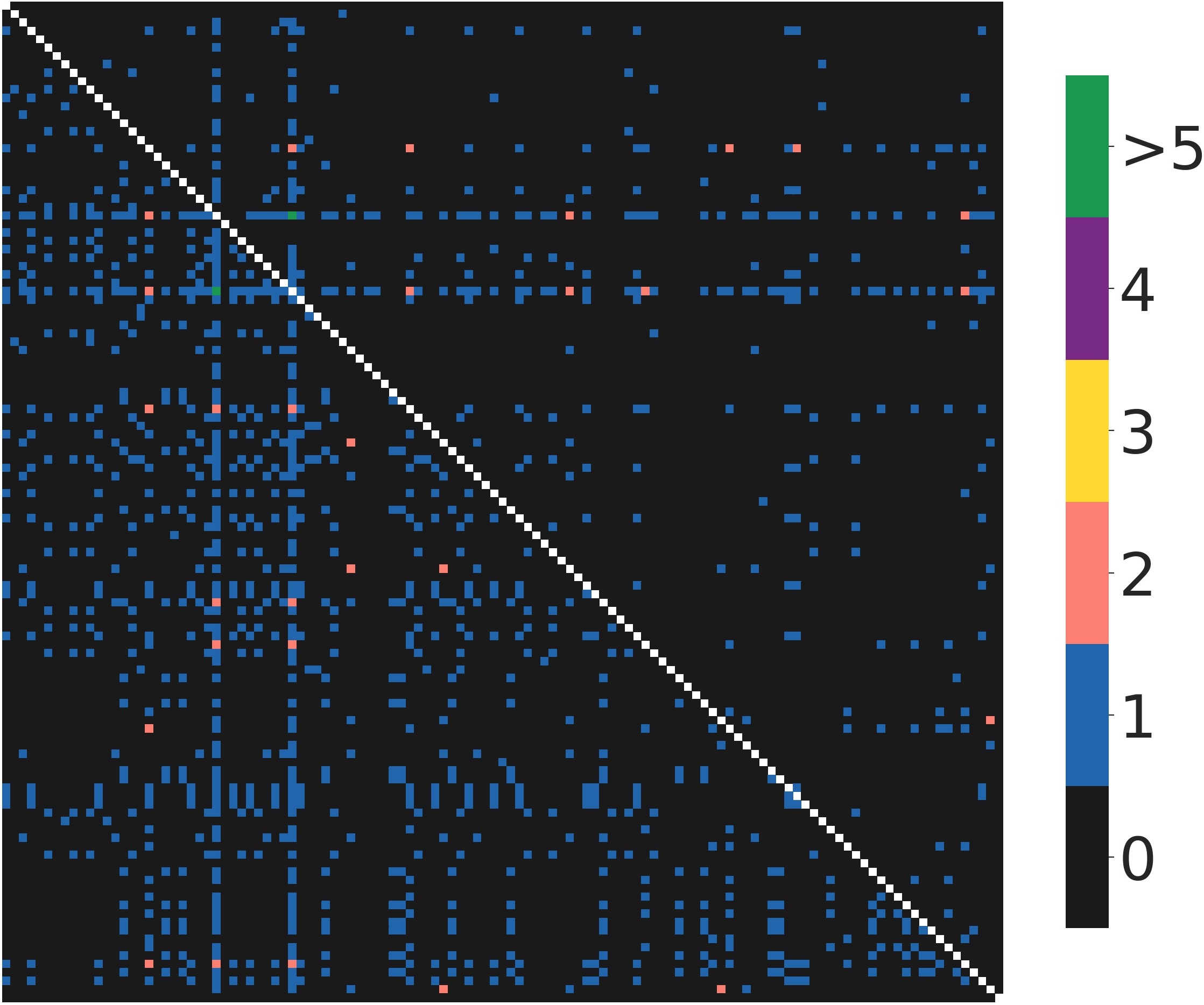}
    \centerline{\scriptsize{(b)}}
  \end{minipage}
  \caption{Overlap of parameter sets across video classes: (a) below diagonal for SPS and above diagonal for VUI (b) below diagonal for PPS and above diagonal for combination of parameter sets. }
  \label{fig:header-model}
\end{figure}

To evaluate the between-class variation, we examined the overlap of headers among video classes.
Figures \ref{fig:header-model}(a) and (b) display corresponding confusion matrices where each row shows the relation between a video class and others in terms of the intersection of the common SPS, VUI, PPS dictionary items as well as their unique combination. 
Due to the symmetric nature of a confusion matrix, in each figure, the area above and below the diagonal is associated with a different parameter set.
In these figures, the intensity at a particular row and column is indicative of the number of overlapping header entries between corresponding video classes.
These confusion matrices essentially demonstrate that encoding settings are not unique to each video class but each class overlaps only with a small number of classes with very few video classes getting confused with a large number of classes. 
This indicates that video classes can be clustered based on their encoding settings.

\section{Proposed Method}
\label{sec:method}

As encoding and encapsulation characteristics of a video file represent two independent aspects, they are complementary in nature. 
Therefore, our method combines both sources of information for more accurate source attribution.
To this end, we first introduce a joint representation for file metadata, named {\em file-metadata tree}. 
Then, we use a two-level hierarchical framework for classification 
as a stratified learning approach will provide better scalability in meeting the challenge of performing attribution in the presence of a large number of source classes.
At the first-level, 
we evaluate topological properties of the file-metadata tree to perform a coarse classification considering three abstractions. 
Then at the second level, we map the file-metadata representation to a feature vector in order to identify the camera model.

\subsection{File-Metadata Tree}

\textcolor{black}{To obtain a coherent representation, the file structure information and encoding parameters are combined in a {\em file-metadata tree} which is represented as a graph $G(V,E,X)$ where $V$ is  node set, $E$ is the edge set showing the relation between nodes, and $X$ is the feature matrix containing a description of each node.
A file-metadata tree consists of two main subtrees.
The first subtree, $G_{c}$, comprises the encapsulation-related metadata with node features indicating box types, field attributes, or attribute values.
In fact, the hierarchical nature of the MP4 file format readily lends itself to such a representation.
Accordingly, the internal nodes of $G_{c}$ correspond to boxes and field attributes reflecting the order they appear in the file.   
The value of each field attribute is represented as a child node descending from the node denoted by the corresponding field attribute.
}

\begin{figure}[htbp]
\centering
    \centering
    \includegraphics[width=0.85\columnwidth]{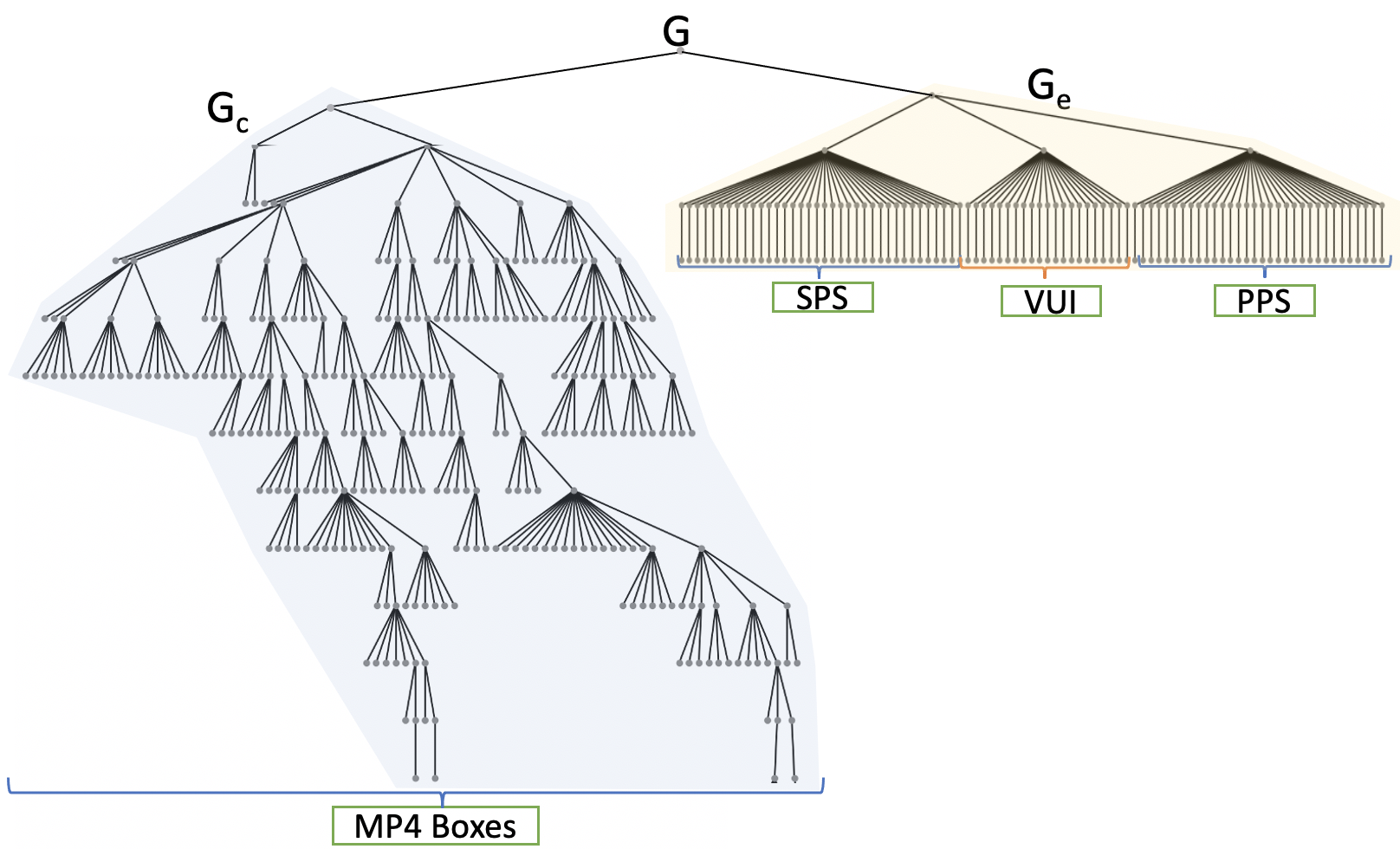}
  \caption{An example file-metadata tree obtained from a video recorded by an Apple iPad2 camera.}
  \label{fig:example}
\end{figure}

\textcolor{black}{
The second subtree, $G_{e}$, comprises the encoding parameters with node features indicating the type of parameter, parameter names, and their values. 
In essence, encoding parameters comprise three, ordered feature vectors of varying lengths.
These parameters are organized into a tree by connecting the SPS, PPS, and VUI to a root node as its child nodes.
Each of these three nodes is then appended as many child nodes as the number of parameters they contain. These second-level nodes are labeled by the name of the corresponding parameters.
Similar to field-attribute values, each parameter node is appended a leaf node containing its value.
The complete file-metadata tree, $G$, is then generated by combining $G_{e}$ and $G_{c}$ as shown in Fig. \ref{fig:example}.}

When extracting file metadata, we used the publicly available software modules.
In the case of H.264 coded files, the encoding parameters are extracted using the H264 BitStream toolbox \cite{h264bitstream}. 
For MP4 files, these include the MP4 Parser library \cite{sanniesm0} and the GPAC library \cite{GPACNigh31} which will be respectively referred to as {\em MP4 Parser} and {\em GPAC}.
Our examination revealed that MP4 Parser yielded some errors when processing MOV formatted files and failed to read some of the fields, such as those including SPS and PPS strings. 
As reported in \cite{xiang2021forensic}, some box fields contain information that reflect a manufacturer-specific usage (such as the \texttt{ilst} box); neither parser supports such non-standard coding of information.    

Nevertheless, we observed that both parsers are able to extract different information and one is not superior over the other.  
Therefore, we used both parsers as part of our method.  
An important consideration here is that a part of extracted file metadata is specific to the contained media and varies from one file to another. 
Previous work has already identified several fields that cannot be used for attribution \cite{yang2020efficient,xiang2021forensic}.
\textcolor{black}{
Those metadata fields are eliminated when creating file-metadata trees.
We additionally excluded fields that carry attributes of encapsulated media data such as track data sizes and all types of entry count values\footnote{\textcolor{black}{Eliminated fields include: {\em TimeToSampleEntry, CompositionOffsetEntry, ChunkEntry, SyncSampleEntry, ChunkOffsetEntry, SampleDependencyEntry, SampleToChunkEntry} and {\em size}.}}.
}

\textcolor{black}{
Ultimately representing video files with file-metadata trees, the video source identification problem can be viewed as a \textit{graph classification} task.
Hence, given a set of labeled file-metadata trees $S$ = \{($G_1$, $l_1$), ($G_2$, $l_2$), \dots, ($G_n$, $l_n$)\}, the goal is to learn a function $f: \mathscr{G} \rightarrow \mathscr{L}$, where $\mathscr{G}$ is the input space of file-metadata trees and $\mathscr{L}$ is the set of video source classes, to predict the label $l$ of an input $G$.
}

\subsection{Learning at First Level}
\label{sec:firstLevelMethod}
The design of our method is mainly motivated by the observation that the topology of a file-metadata tree does not vary significantly across 
camera models and video processing software and, in fact, remains very similar or the same for many classes of videos. 
Hence by first identifying metaclasses that group together several video classes, this step facilitates better learning of the nuances in variation of field and parameter values across similar video classes. 
In a way this can be viewed as analogous to performing brand identification first.
This approach, however, further exploits the fact that different brands or editing tools may exhibit similar encapsulation and encoding behavior.
\textcolor{black}{
To capture this behavior, the leaf nodes of $G$, containing values of box fields and encoding parameters, are removed. Denoting this reduced height tree by $G_r\in\mathscr{G}$, learning at this level can be formulated as identifying a function $m: \mathscr{G} \rightarrow \mathscr{H}$, where $\mathscr{H}$ is a set of metaclasses.
In essence, $m$ categorizes file-metadata trees into metaclasses in an unsupervised fashion, i.e., $m(G_r)=h\in\mathscr{H}$. 
We consider three different constructions for $m$ depending on how node features are utilized and whether node ordering information is incorporated.
}

\subsubsection{Tree Topology Embedding}
\textcolor{black}{This construction of $m$ completely disregards node features $X$ to perform a coarse characterization.}
That is, only the shape of a file-metadata tree is taken into consideration to detect different structures of trees.
For this, we treat the file-metadata tree as an undirected graph and use a graph representation learning approach.
A graph embedding essentially transforms a whole graph into a low-dimensional vector to allow for the classification of graphs. We considered several graph embedding techniques for our task \cite{cai2018simple,schulz2019necessity,tsitsulin2020just}\footnote{\label{karateNote}All graph embeddings are created by the {\em Karate Club} \cite{karateclub} software package using their default hyperparameter settings of the 1.2 release.}. 
Our tests revealed that these approaches perform very similarly. 
(We defer further discussion of this to Sec. \ref{sec:discussion}.)  
Based on our evaluation, we selected the {\em local degree profile} (LDP) \cite{cai2018simple} representation scheme which yields a fixed embedding for the same graph, thereby avoiding the need for error-prone clustering. 
This representation essentially characterizes each node and its 1-hop neighbors through five degree statistics.
The empirical distributions obtained by aggregating these statistics over all nodes in a graph are then concatenated into a feature vector.
This resulting representation is used as a basis for differentiating file-metadata trees, where each distinct embedding is treated as a separate metaclass.

\subsubsection{Tree Embedding}
A more complete characterization requires also incorporating the labels of nodes in a tree.
For this, the node features $X$ are generated using the textual node labels (i.e., box and field attribute names).
To better capture the relationship between node labels a distributed vector representation based on the Word2Vec \cite{word2vec} CBOW model is trained by treating each root-to-leaf path comprising a sequence of node labels as a sentence\footnote{The Word2Vec representations are generated using default settings of Gensim \cite{gensim} implementation by setting the embedding size to 10.}.
To obtain graph representations, we considered three graph embedding methods that can utilize node features $X$, including FEATHER-G \cite{feather}, GeoScattering \cite{gao2019geometric}, and Graph2Vec \cite{narayanan2017graph2vec}\footnotemark[2].
These graph embeddings, however, yield slightly different representations for the same graph.
Therefore, resulting embeddings need to be clustered together to identify metaclasses. 
For this purpose, the K-means clustering method is used, and the number of clusters is decided by using the Elbow method and Silhouette score.
In our evaluation, we determined that GeoScattering embeddings perform slightly better.
(Related aspects will be further discussed in Sec. \ref{sec:discussion}.)

\subsubsection{Exact Tree Match}
The last construction of $m$ also takes into account the order of nodes in addition to tree topology and node labels by performing exact tree match. 
Essentially, the order of boxes in MP4-like files can be captured by invoking a depth-first search (DFS) from the leftmost node of the corresponding file-metadata tree. 
To exploit this, DFS-ordered node labels obtained from \textcolor{black}{$G_r$} are organized as a sequence and cryptographically hashed to produce a value.
Each hash value is then treated as an index to a separate metaclass.

In all the three settings, an important concern is the comprehensive coverage of metaclasses.
Since some of the video classes have fewer video instances than others, it is likely that during training our method will not generate descriptors for all metaclasses.
As a result of this incomplete view, corresponding metaclasses for some video files may not be identified.
To account for such cases, we incorporate an {\em unknown} metaclass to each clustering setting at the first level. 
Hence during testing, if a file-metadata tree cannot be structurally matched to any of the known metaclasses, it is assigned to the {\em unknown} class.

\subsection{Classification at Second Level}

After file-metadata trees are grouped into metaclasses based on their structural properties, a more refined classification is performed. 
At this level, feature and parameter values are used to build separate machine learning models for each metaclass.
\textcolor{black}{That is, for a given set of attributed file-metadata trees $S$ = \{$(G_1,h_1,l_1)$, $(G_2,h_2,l_2)$, \dots $(G_n,h_n,l_n)$\}, where $h_i=m(G_{r_i})$ is determined at the first level, we generate functions $f_h$ : $\mathscr{G} \rightarrow \mathscr{L}$, $\forall h \in \mathscr{H}$.}
This again requires mapping the information contained in file-metadata trees to a feature representation. 
For this, we use the track and type-aware representation introduced in \cite{xiang2021forensic} that creates a feature vector with each dimension holding occurrence frequency of root-to-node path descriptions and values of non-categorical values in the file-metadata tree as it yields more compact feature vectors.
\textcolor{black}{Then for each metaclass $h$,  a decision tree model is used to learn the function $f_h$ that maps the resulting feature vectors to their source-class labels.}

\subsection{Tackling Partial Video Files}
When a video file is partially available, a condition encountered frequently during recovery of data from disks, memory, and SSDs using file carving tools, file metadata will be largely inaccessible.
Especially for file formats that are mainly designed for video playback (as opposed to streaming), this is a major problem because metadata information comprise only a very small part of the overall file size, and they may not be located during file carving.

To make such video file fragments playable, \cite{altinisik2021automatic} introduced a method for estimating encoding parameters through a search over parameter space by incorporating decoding error messages\footnote{\textcolor{black}{The tool is available at: https://github.com/FileScraper/tool }}. 
Although the space of parameter values defined by the H.264 standard is extremely large, many parameters vary slightly across camera models and for several parameters choosing the highest or lowest categorical setting will not curtail decoding.
This work essentially showed that a large number of videos can be decoded without having access to actual parameters by only determining a core set of 10 parameters in the SPS and PPS. 
Crucially, five of these parameters relate to how the bitstream is parsed and the rest determine the entropy coding mode, frame width, and base-quantization level.

Given this capability, we also investigate how reliably the source class of a video can be identified using a subset of coding parameters.
With this objective, we apply the method in \cite{altinisik2021automatic} to coded frame data extracted from videos to estimate their coding parameters with one modification. 
In its operation, this method does not identify frame height as a core parameter because an incorrect setting causes the decoded frame to be either truncated or elongated through content repetition. 
To improve the distinguishability, we also incorporate the frame height information. 
For this, we changed the default value for frame height to the largest possible value which inadvertently caused interpolation of the last line of $8 \times 8$ picture blocks.The height is then estimated by determining the repetition point through the correlation of block content.  
The resulting 11 parameter values are then treated as a feature vector and used for model building directly.

\section{Results}
\label{sec:results}

We now examine the effectiveness of the proposed source identification method.
Our evaluation assumes a closed-set identification scenario.
For this, we merged together VISION, ACID, SOCRatES, and EVA-7K datasets to obtain a large pool of video data.
The former three datasets include videos captured by smartphones and digital cameras whereas the last one includes videos of various provenance downloaded from different social media platforms and edited by several tools\footnote{In disagreement with its description, the SOCRatES dataset did not include videos captured by the following models: Wiko Rainbow Up 4G, LG G4, Apple iPhone 5, Sony Xperia E3, and Meizu M3 Note.  In addition, videos associated with the Motorola Moto G could not be played. The dataset, however, included another instance of the same model.\\
We also excluded Adobe Premiere and ExifTool videos in the EVA-7K dataset from our combined dataset as both editing tools hardcode source information within the Extensible Metadata Platform (XMP) headers.}.
This resulted in 20,153 videos from 112 camera models of over 28 brands, 3 editing tools, and 4 social media platforms with 10-1,400 videos per video class.
Our examination showed that except for those associated with eight camera models, all videos are captured in H.264-MP4 format.
Out of these camera models, five models used H.264 encoding, one used the MP4 file container format, and the remaining two used neither. Since the resulting dataset is significantly unbalanced, in our performance evaluation we used the balanced-accuracy metric which evaluates the success of distinguishing each camera model at an equal footing.

\subsection{Comparison of Coding and Encapsulation Characteristics}
We first quantify the level of distinguishability provided by encoding and encapsulation characteristics individually and together. 
In our evaluation, we use a flat classification approach similar to that proposed by earlier work \cite{yang2020efficient,xiang2021forensic}.
For this, we consider two file-metadata tree representations to determine the achievable identification accuracy, which will be used as a baseline. 
The first is the sparse vector representation based on root-to-leaf path descriptions that combines two feature vectors \cite{yang2020efficient}, one obtained using an unordered list of path entries with only field names and the other using an unordered list including both fields and values (see Fig. \ref{fig:represent}). 
The second is the track and (field) type-aware representation introduced in \cite{xiang2021forensic} without feature selection and LDA-based feature dimension reduction.
When using this representation all codec parameters are treated as non-categorical variables.
To also assess the impact of metadata parsers, file-metadata trees are generated for all the videos using two public parser libraries \cite{sanniesm0,GPACNigh31}.
Feature vectors are then used to train a decision tree (DT) classifier with balanced class weights\footnote{We also considered other classification models in our tests, we determined that DT and Random Forest classifiers perform similarly and much better than KNN, SVM, and XGBoost classifiers.}. 

Table \ref{tab:flatClass} provides balanced accuracy values obtained after five-fold cross validation with 100 runs to reduce any sampling bias.
The first column denotes the library used for parsing MP4 metadata, the second one shows the method used for turning file-metadata trees  into feature vectors, and the third column displays the source of metadata, i.e., whether based on container, codec, or the combination of the two. The balanced accuracy results obtained in identifying 119 video classes involving 20,153 videos are given in the fourth column.
Tests are also repeated considering only MP4 formatted videos, which reduced our combined dataset to 112 classes with 18,456 videos, as presented in the last column.

Results show that when compared in isolation, codec parameters provide less distinguishability (line 9) than MP4 file metadata (lines 1, 3, 5, and 7).  
This can be attributed to the lower dimensional nature of codec parameters compared to the rich information contained in the MP4 file metadata.
However, when combined together, the complementary nature of both information sources is revealed as in all test settings combined features yielded the best accuracy (lines 2, 4, 6, and 8).
More specifically, the highest balanced accuracy achieved by either sparse \cite{yang2020efficient} or track and type-aware \cite{xiang2021forensic} feature representation of container characteristics in the overall dataset is 84.3\% (line 5).
Whereas in the combined setting, balanced accuracy reaches 90.22\% (line 4) with an improvement of 6\% in accuracy.
Even when limiting the analysis to only MP4 formatted videos, classification accuracy improves by 1.4\%, from  89.02\% (line 5) to 90.42\% (line 4).
Overall, these results indicate that the encoding aspect of a video file provides improved discrimination capability.  

Our findings also demonstrate that the more compact track and type-aware feature representation, in general, yields better performance than the sparse representation.
Further, results show that the metadata extracted by both parsers are indeed different.
Comparatively, the GPAC parser is more effective for both feature representations. 
In the combined setting, however, MP4 Parser is found to yield the best performance. 
One factor contributing to this result may be the incorporation of SPS and PPS strings extracted by
the GPAC parser (which are disregarded by MP4 Parser) with individual parameter values which might be confusing the DT classifier.

\begin{table}[!ht]
	\centering
	\caption{Video-Class Identification Accuracy for Flat Classification}
	\label{tab:flatClass}
	\resizebox{\linewidth}{!}{
    	\begin{tabular}{r|c|c|c|c|c|}
    	\cline{2-6}
                   & & \multicolumn{2}{c|}{Feature}                      &All Videos   & MP4 Videos  \\ 
    	& Parser     & Representation            & Source     &(119 classes)  &  (112 classes)  \\\cline{2-6}
    	1&\multirow{4}{*}{\parbox{1cm}{\centering MP4 Parser \cite{sanniesm0}}} 
    	           & \multirow{2}{*}{\parbox{1cm}{\centering Sparse}}     
    	                                       & Container  & 77.84      &  82.02   \\ \cline{4-6}
    	2&           &                           & Combined   &87.43      &  87.96   \\ \cline{3-6}
        3&	       & \multirow{2}{*}{\parbox{1.5cm}{\centering Track \& Type Aware}}    
        	                                   & Container  &82.65     &  87.19   \\ \cline{4-6}
    	4&           &                           & Combined   &\textbf{90.22}  &  \textbf{90.42}    \\ \cline{2-6}
    	5&\multirow{4}{*}{\parbox{1cm}{\centering GPAC \cite{GPACNigh31}}}     
    	           & \multirow{2}{*}{\parbox{1cm}{\centering Sparse}}      
    	                                       & Container  &84.30     &  89.02  \\ \cline{4-6}
    	6&           &                           & Combined   &88.62     &  89.10   \\ \cline{3-6}
        7&	       & \multirow{2}{*}{\parbox{1.5cm}{\centering Track \& Type Aware}}     
        	                                   & Container  &84.10    &  88.78   \\ \cline{4-6}
    	8&           &                           & Combined  &88.64   &  88.80  \\ \cline{2-6} 
\cline{2-6}
    	9&\multicolumn{2}{c|}{}                 & Codec      & 63.84   &  64.27\\ \cline{2-6}                                         
    	\end{tabular}
	}
\end{table}

\subsection{Two-Level Hierarchical Classification Method}
\label{sec:metod-results}

At the first level, the hierarchical method groups file-metadata trees into metaclasses based on high-level structural properties.  
In this regard, the LDP graph embedding of file-metadata trees results in a 1280-bit 
representation, the node-labeled graph embedding results in a 3K-bit representation, and the tree hashing approach yields 256-bit index values. At the second level, file-metadata trees in each metaclass are mapped to feature vectors using the track and type-aware representation, and a DT classifier is built for each metaclass. 
During training, each class is weighted based on their weights in the overall training set. 
To exploit the advantages of both parsers, we used GPAC at the first level and MP4 Parser at the second level
as the latter performed slightly better in flat classification.
While testing the system, in cases where a metaclass cannot be identified for a given file-metadata tree, it is assigned to the  unknown class.
Those videos are then classified using the best model in Table {\ref{tab:flatClass}} (line 4).

Table \ref{tab:2levhie} presents five-fold cross validation results averaged over 50
randomizations of the training and testing datasets (lines 1-3). That is, in each run, we used 80\% of the videos to identify metaclasses and to train several DT classifiers, while the remaining 20\% were used to measure the balanced identification accuracy of the overall system.
The first column of the table shows the method used for clustering file-metadata trees in the first layer, and the following three columns 
present results of the first-level classification obtained on the whole dataset.
The total number of metaclasses generated by each abstraction is given in the second column. 
The relatively low number of metaclasses created by the three abstractions validates our intuition that file-metadata trees are in fact structurally very similar and do not exhibit strong variation among videos.
The percentage of file-metadata trees assigned to the unknown class (third column) further verifies that structures associated with each metaclass
are quite invariant. 
It is determined that except for the second abstraction, on average only 0.1\% of videos are assigned to the unknown class.
(When clustering GeoScattering embeddings, we used an Euclidean distance of 3.5 as a threshold for assigning a video to the unknown class.)
The error in assigning a file-metadata tree to an incorrect metaclass at the first level is given in the fourth column, and it is found to be  extremely small in all cases.

Video source identification accuracy values obtained at the second level using the three abstractions are given in the last two columns considering both the whole dataset (119 classes) and its reduced version (112 classes).
Overall, results show that the three abstractions yield similar performance, but the use of tree-hashing at the first level yields the best performance. 
\textcolor{black}{Compared to the best results obtained using the flat classification of container features (as presented in line 4), our proposed method provides an additional 6.5\% improvement in balanced accuracy, from 84.3\% to 90.8\%, on the whole dataset.
When only MP4 videos are considered, the improvement is measured to be 2.1\%, increasing from 89.02\% to 91.14\%.}

\begin{table}[!ht]
	\centering
	\caption{Source Identification Accuracy with Two-Level Hierarchical Classifier}
	\label{tab:2levhie}
	\resizebox{\linewidth}{!}{
    	\begin{tabular}{r|c|c|c|c|c|c|}
    	\cline{2-7}

    	&\multicolumn{6}{c|}{ }\\[-6.5pt] 
    	&\multicolumn{6}{c|}{{\bf Two-Level Hierarchical Classification Method}} \\ 
    	&\multicolumn{6}{c|}{ }\\[-6.5pt] 
    	\cline{2-7}

        & \multicolumn{4}{c|}{First Level}  & \multicolumn{2}{c|}{Second Level} \\     	&Clustering             & \# of   & \% of videos in  & Incorrect              & \multicolumn{2}{c|}{Classification Results}  \\
    	&Method                     & metaclasses   & {\em Unknown} metaclass & classification (\%)  & All           &   MP4    \\ \cline{2-7}
    	1&LDP              &   164   &   0.09       &   0.03                  & 90.74         &   91.0  \\ \cline{2-7}
    	2&GeoScattering             &   180   &   2.65       &    0.31                 & 90.81         &  90.89 \\ \cline{2-7}
    	3&Tree hashing              &   178   &   0.11       &   0.02       & \textbf{90.84}      &  \textbf{91.14}\\ \cline{2-7}
        
        \multicolumn{7}{c}{ }\\[-6pt] 
        \cline{2-7}
        
        &\multicolumn{6}{c|}{ }\\[-6.5pt] 
        &\multicolumn{6}{c|}{{\bf Flat Classification}} \\ 
        &\multicolumn{6}{c|}{ }\\[-6.5pt] 
        \cline{2-7}
        
        4&Only Container            &  -     &      -        &         -                &84.30        &   89.02    \\\cline{2-7}

    	\end{tabular}
	}
\end{table}

\textcolor{black}{
We also evaluated our method by individually training and testing a model on each dataset.
Table \ref{tab:diffDataset} provides the corresponding identification accuracy values obtained for VISION, ACID, SOCRatES, and EVA-7K datasets.
Here, EVA-7K dataset is further divided into two parts to better capture the influence of editing tools and social media platforms on file metadata due to re-encoding and re-encapsulation of videos.
These results show that among the four datasets  EVA-7K and ACID pose the least difficulty to video-class identification.
In contrast, SOCRatES is found to be the most challenging dataset.
This can be attributed to the fact that this dataset includes 56 video classes out of which 38 comprise only 10 video samples with an overall average of 16 samples per class.}

\begin{table}[!ht]
	\centering
	\caption{Source Identification Accuracy with Two-Level Hierarchical Classifier on Each Dataset }
	\label{tab:diffDataset}
	    	\begin{tabular}{r|c|c|c|}
    	\cline{2-4}    
    	  & Dataset       & \multicolumn{2}{c|}{Proposed (Table \ref{tab:2levhie} line 3)}\\ 
    	  &               & All Videos & MP4 Videos               \\ \cline{2-4} 
    	1 & VISION        & 90.68      & 90.64                    \\ \cline{2-4} 
    	2 & ACID          & 94.41      & 96.65                    \\ \cline{2-4} 
    	3 & SOCRatES      & 87.52      & 87.32                    \\ \cline{2-4} 
    	4 & EVA-7K (Social Media Platforms)  & 100        & 100                      \\ \cline{2-4}   
    	5 & EVA-7K (Editing Tools)     & 99.84      & 99.84                    \\ \cline{2-4}   
      	6 & \textbf{Overall}         & \textbf{90.84}      & \textbf{91.14 }                   \\ \cline{2-4}
  	
    	\end{tabular}
	\end{table}

\subsection{\textcolor{black}{Scalability Assessment}}

\textcolor{black}{
An important question concerning the use of our source attribution method is its scalability under a large number of source classes.
To partially explore this question, we created another dataset. 
For this, we obtained 92,603 videos from the public video sharing website \texttt{lbry.com} which, 
unlike other popular video sharing websites, does not by default transcode and re-encapsulate uploaded user videos. 
Videos were obtained from 14,916 platform user accounts by iteratively crawling the suggested video links on the main webpage.
It can plausibly be assumed that this collection is of more diverse provenance than the combined test video set of 20K videos.
An important limitation of this video set, however, is that source-class labels (i.e., camera model and/or processing software suite) are not available.
Therefore, attribution accuracy cannot be validated on it. 
In the absence of this information, we alternatively examined the number of metaclasses encountered at the first level of our learning hierarchy to determine its scaling behavior.
We also studied the increase in the number of MP4 file container features used for classification at the second level.}

\textcolor{black}{
In our analysis, we first extracted file-metadata trees from all videos and grouped them into metaclasses using the tree hashing-based abstraction approach.  
We determined that this yielded a total of 2,027 metaclasses, as opposed to 178 seen in the earlier tests.
The 10 largest metaclasses are found to include close to 71.5K videos with the largest one including around 37K videos. 
We speculate this cluster of videos is likely to be those transcoded by the platform (at the request of the user) as it included videos of 6,173 users.
Each of the remaining metaclasses included 1-5K videos showing that certain encoding and encapsulation behaviors are more prevalent than others. 
On average, 2.09 hashes per user are generated with 8,019 users yielding only one (but not necessarily unique) hash. 
This shows that within-user variation in encoding and encapsulation characteristics is low.
We also identified that 352 metaclasses are associated with videos of specific users, potentially indicating some rarely used camera-models, editing tools, or settings. 
}

\textcolor{black}{
The change in the dimensionality of the sparse and the track and type aware feature representations used for classification is another concern. 
We determined that depending on the choice of the parser and the representation, this video set yields feature vectors of size 35K-70K.
Considering feature vectors obtained from videos in the combined test dataset (5K-11K), this constitutes around seven-fold increase in dimension.
This can be attributed to a larger number of video source classes included in this collection of videos. 
An implication of this observation is that a flat classification scheme will get computationally less feasible for much larger datasets.
This further signifies the need for a hierarchical classification approach.
}

\subsection{Comparison with Other Methods}

We compare our proposed method with the approaches of three methods that utilize file metadata for video source identification  \cite{yang2020efficient, gelbing2021video, xiang2021forensic}.
Since public implementations for all these methods were not available, we implemented them to the best of our ability and applied them to our context.
In \cite{yang2020efficient}, Yang et al. focused on the video integrity verification problem while proposing an improvement over \cite{iuliani2018video} which used the likelihood ratio of ordered root-to-leaf paths (including field-value pairs) to perform brand identification and integrity verification.
This work reported that the use of the sparse feature representation obtained by combining the above path entries with unordered root-to-leaf paths with only field names in conjunction with a DT classifier yields higher accuracy.
This method uses the public MP4 Parser library, and its feature extractors are available. 
In this regard, the results given in the first line of Table \ref{tab:flatClass} (77.8\% and 82.0\%) correspond to achievable accuracy by this method.

Gelbing et al. \cite{gelbing2021video} introduced the insertion/deletion resistant ordering of root-to-node path elements and used a weighting-based matching to evaluate the similarity of file metadata across brands and unique devices.
Accordingly, the weight for each path entry is determined based on the extent to which they contribute to the identification of each model.
The authors also proposed blocking a list of boxes because they cannot be used for attribution.   
Our implementation, excluding those boxes, yielded an area under the curve (AUC) of 91.1\% for device identification on the VISION dataset which is very close to the 90.3\% value reported in the paper.
However, the computation of the AUC metric allows matching a video file to multiple video classes, thereby creating many correct decisions.
When only the best-matching class is considered, for the same implementation, an average top-1 similarity score of 66.1\%  over 100 runs is obtained in identifying 30 camera models in the VISION dataset.
In comparison, our proposed method yielded a balanced accuracy of 90.6\% on the VISION dataset.
We also incorporated the feature mapping introduced in \cite{gelbing2021video} to the two-level hierarchical classification method, and obtained a balanced accuracy of 90.5\% on the VISION dataset which was lower than the two representations in Table \ref{tab:flatClass}.

In \cite{xiang2021forensic}, Xiang et al. introduced another feature representation using a custom parser that can also extract information in \texttt{ilst} boxes and XMP headers.
This method deploys a clustering-based feature selection approach and performs LDA to obtain a two-dimensional feature followed by nearest-neighbor classification.
When implementing this method, we used MP4 Parser to obtain the track and type-aware feature vector as it yielded better results than the GPAC parser.
To determine the impact of using the public parser (as opposed to the deployed custom parser) on performance, we validated our implementation on a closed-set brand-attribution setting involving the VISION dataset as considered in the paper.
This yielded an identification accuracy of 99.77\% which is very close to the 99.8\%\footnote{It is reported that in this test scenario the method achieves 99\% accuracy in identifying two brands while getting 100\% accuracy on all others. This corresponds to a balanced brand classification accuracy of 99.8\%.} value found in the paper.
When performing feature selection and LDA, we considered all clusters of size up to 3K features which at the best case yielded a balanced accuracy of 74.5\% on the overall dataset.
We believe reducing the feature vector size drastically may not be appropriate when the number of video classes is large. 
Therefore, we used track and type-aware feature vectors as part of a DT classifier, without dimension reduction, which resulted in 84.1\% balanced accuracy as given in Table \ref{tab:flatClass} (line 7) implying that a DT is more effective in identifying critical features.

Overall these results demonstrate that the incorporation of encoding parameters with the conventional
MP4 based file metadata in a multi-level classification approach improves the achievable video source identification accuracy by almost 6.5\%.

\subsection{Exclusion of User Adjustable Settings}
Many cameras as well as camera apps of smartphones allow users to choose among a set of resolutions, frames-per-second, and a compression quality when capturing a video. 
Since these are user set parameters, comprehensive coverage of all possible settings associated with a camera model may not be possible. 
Therefore, we investigate the extent to which these settings affect the identification accuracy.
For this purpose 7 parameters in SPS and 24 fields in MP4 boxes are excluded when mapping file-metadata trees to feature vectors.
Repeating the same training and test procedure described in Sec. \ref{sec:metod-results} 
corresponding identification accuracy values in the third row of Table \ref{tab:2levhie} are, respectively, obtained as 
90.85\% and 91.10\%.
The fact that almost the same  performance could be obtained indicates that video file metadata contains rich model-specific information and that this diversity can compensate for the lack of considered user adjustable settings.

\subsection{Partial Video Files}
This test setting concerns the extreme case where only a minimal subset of encoding parameters are available for source identification. 
For this, the automatic header generation method of \cite{altinisik2021automatic} is used to estimate the values of the 10 core parameters along with the picture height from original (unedited) videos captured by a camera.
We must note here that changing the values of encoding parameters does not necessarily cause a decoding failure. 
As an example, a video encoded with the frame {\em cropping flag} set to 0 can still be decoded when the {\em cropping offset} is set to 0 while the cropping flag is set to 1.
Therefore, the estimated parameters that allow decoding of the video data may not necessarily be the same as those in the original file.
In fact, 6 of the 11 parameters, including the base quantization value, cropping (flag and offset), picture order count, number of 
frames, and height, can potentially take alternative values without a failure in decoding.

We used the first coded frame of 12,892 H.264 video captured by 109 camera models to estimate the 11 parameters that allowed successful decoding of the picture.
The comparison of estimated parameter values from all videos with their original values revealed that around 80\% of cropping and quantization parameters are the same as their original values.
Similarly, around 70\% of the values corresponding to picture order count and the number of frames parameters were found to be the same.
In the case of picture height, the average error across all videos is measured to be 34.5 pixels with 61\% of the height values having an error less than 10 pixels. 

These estimated parameters are then used as feature vectors to build a DT classifier which resulted in a balanced accuracy of 57.2\%.
Although this marks a considerable drop from the results of Table \ref{tab:2levhie} (from 91\% to 57\%), it is surprising to determine that only 11 estimated parameters can achieve this level of discrimination among 109 camera-models. 
To further test the distinctiveness of these parameter values, we developed another classifier for brand-level identification by combining together videos of all models of a brand as a single class.  
For this task, the achievable accuracy is found to be 81\%.
Similar to the previous setting, we also performed the test by disregarding four user-adjustable parameters that relate to frame height and width.
As expected, this caused a further decrease in performance, and 39.7\% and 69.2\% balanced accuracy results were obtained for both scenarios.

\section{Discussion}
\label{sec:discussion}

In the presence of thousands of camera models and many editing tools, the most important challenge for video source identification methods is scalability.  
This requires a comprehensive approach that combines several sources of information, including those obtained from photos and videos through both content and file-metadata analysis. 
Camera-model identification has long been the goal of content analysis methods, more specifically in the context of photographic images.
Accordingly, the highest performance has so far been reported by data-driven methods that use deep learning frameworks to 
learn features or as part of an end-to-end classification approach \cite{bondi2016first,marra2017study,guera2018reliability, rafi2019application}.
In this respect, when tested on images of 54 camera models in DRESDEN and VISION image datasets, the method of \cite{marra2017study} 
achieved 83\% classification accuracy.
In comparison to these results, our findings show that file-metadata based video source identification provides an important complement to existing content-based analysis methods.

At its core, our approach views encoding and encapsulation as two complementary and independent aspects.
Therefore, other representations for container-based file metadata can be easily incorporated with the codec parameters to perform the 
hierarchical classification. 
Similarly, the use of more advanced parsers that can extract non-standard, manufacturer-specific information will further improve the reported accuracy values as our method builds on these capabilities.

{\bf Application of neural network models.} As an alternative to our current approach, we also considered recurrent neural networks to exploit the sequential nature of the file metadata. Since in its original form, file metadata appears as a sequence of boxes along with their fields and values, we also applied  GRU/LSTM \cite{lstm,cho2014learning} models by posing the problem as a sequence classification task.
Our results showed that these sequential models could not sufficiently capture the inherent tree-structure of the data and yield inferior results.
This may simply be attributed to the limited amount of video data available in some video classes (10 videos), and the effectiveness of the DT classifier in identifying clear decision boundaries for our mostly categorical features.

We further tested some of the unsupervised graph learning methods on our dataset. 
These whole graph embedding approaches allow embedding of file-metadata trees (including all branch nodes and leaf nodes containing values) to a vector space which can then be used for creating a classification model. 
In this regard, the well-known Graph2Vec \cite{narayanan2017graph2vec} and GL2Vec \cite{chen2019gl2vec} graph representations, coupled with the KNN classifier, achieved relatively low balanced accuracy values of around 67\%.
In contrast, more recently introduced FEATHER-G and GeoScattering graph embedding methods achieved a performance slightly above 85\%.
This finding indicates that in the presence of larger datasets, graph learning methods have strong applicability to the model-identification task.

We also examined the distinctiveness inherent to the shape of file-metadata trees by disregarding the node labels and values.
For this purpose, we performed topology based classification of file-metadata trees using LDP \cite{cai2018simple}, NOG\cite{schulz2019necessity}, Slaq-VNGE, and Slaq-LSD \cite{tsitsulin2020just} graph embeddings which overall yielded balanced accuracy values in the range of 45-56\% with LDP and Slaq performing the best.

{\bf Best achievable performance on our dataset.}
To better understand the room for improvement on our dataset while using the two parsing tools and codec parameters, we compared the extracted metadata from all videos in a pair-wise manner. 
This side-by-side comparison excluded any metadata related to video content that were eliminated during previous tests. 
Over 119 video classes, we identified 10 tuples of video classes, including nine pairs and one triplet, that have exactly the same features.
This essentially indicates that a confusion among 21 video classes is unavoidable, and only 10 of these video classes can be correctly distinguished. 
In other words, when all remaining videos are matched correctly to their classes, the best achievable balanced accuracy will be limited to 90.8\% with only 108 out of the 119 classes being identified accurately.
In the case of MP4 formatted videos, the best achievable accuracy would be 91.1\% due to collision among 10 video classes out of the 112. 
Overall, these results show that our proposed method already achieves a perfect discrimination on this dataset.
It can be noticed that our accuracy values are marginally better than these performance bounds. The slight difference of 0.04\% is essentially due to the randomness in how data is partitioned into folds and the resulting minute variations in class weights which cause
different classes of each confused tuple to be selected in different folds.

\textcolor{black}{
{\bf How to assign labels: camera model vs. firmware.}
When performing source attribution, we used source camera model of each video as its class label.
The firmware, i.e., the software that runs a digital camera, may alternatively be used for labeling video source classes.
In fact, a camera’s firmware may be updated several times over its lifetime. Further, different models of a brand may share the same firmware.
This, however, does not necessarily mean that two different firmware will perform encoding and encapsulation distinctively. To determine how camera model and firmware level attribution compare, we performed further analysis.}  

\textcolor{black}{
Among the four public datasets used in our study, device firmwares are available only in the VISION dataset. 
Therefore, we focus on the subset of video classes contained in this dataset.  
Examining the incorrectly classified videos associated with 21 video classes (out of the 119 classes), we determined that eight are part of the VISION dataset. 
By comparing their firmware, we determined that only in the case of two OnePlus models (A3000 and A3003) the confusions can be explained by the matching camera firmware. 
In contrast, two Huawei models (P9 and P9 Lite), two iPhone models (6 and 6 Plus), and two Samsung models (Galaxy S3 Mini and S3 Mini GT) are found to yield, at the group level, the same file-metadata tree despite running on different firmware. 
As these models are contemporaries of each other, we believe the indistinguishability of these models is still due to their firmware exhibiting the same encoding and encapsulation behavior. 
}

\textcolor{black}{
We also repeated the same source attribution test by rearranging videos captured by 30 camera models on the basis of their firmware, rather than the source camera model.
This yielded 26 firmware classes, and the corresponding balanced accuracy is found to slightly decrease from 90.6\% (as given in Table \ref{tab:diffDataset}) to 89.5\%.
To better understand the underlying phenomenon, we examined whether a camera of a particular model and firmware can be distinguished from all other cameras either with the same model or the firmware. 
For this, we identified three cameras, an iPhone 4 and two iPhone 4S models, where the two iPhone 4S models have different firmware and the iPhone 4 model uses the same firmware as one of the 4S models. 
Under both test settings, i.e., model- and firmware-level attribution, these three models could be perfectly distinguished from each other.
This essentially indicates that file-metadata characteristics of a video may vary depending  on both model and firmware designation of a camera.}

{\bf Robustness issue.} Another important issue is the robustness of file-metadata-based characteristics to video editing or tampering.
Essentially, any video file can be repackaged in one of several video file formats without ever modifying the video content, thereby removing 
the majority of encapsulation-related characteristics. 
In this regard, results of \cite{quinlan1986induction,yang2020efficient,xiang2021forensic} indicate that existing tools and systems do not exhibit such behavior, and in most cases preserve most of the metadata in the original file after editing.
Ultimately, however, modifying the encoding parameters will require re-encoding of the original video.
That is, a change in the encoding parameters will introduce double-compression artifacts which can possibly be identified through content-based analysis methods.
This essentially provides a mechanism to determine whether a given set of encoding parameters should be used for source identification.

\section{Conclusion}
\label{sec:conclusion}
In this work, we introduce a new video source identification method that exploits encapsulation and coding characteristics of video files.
The novelty of our method is twofold. 
First, we consider the use of H.264 codec parameters and contrast their distinctiveness with the organization and content of the boxes used in MP4-like files.
Second, we combine the two characteristics in a file-metadata tree to obtain a joint representation of a video file.
The resulting representations are then used to build a classification model.
Identification accuracy values measured on a video set obtained by combining four public datasets show that the proposed method can achieve 91\% accuracy in distinguishing 119 video classes.
Our results establish that video file metadata strongly characterizes the model of camera that captured it. 
The next steps for research in this area must apply and adapt our approach to a larger number of camera models and combine it with content-based analysis methods to develop a more comprehensive approach to distinguish the camera model of a video.

\appendix
\textcolor{black}{The encoding parameters that exhibit within-class variability across 119 video classes are listed in Table \ref{EncodingId}. The index for each parameter corresponds to Parameter ID given in the x-axis of Fig. \ref{fig:parameterLevel}. For parameter descriptions, we refer the reader to \cite{tsbmail}.} 

\begin{table*}[t]
	\centering
	\caption{\textcolor{black}{The mapping between parameter IDs and names} }	\label{EncodingId}
	
	\begin{tabular}{|c|c||c|c||c|c|}
		\hline
	ID & Parameter  & ID & Parameter& ID & Parameter \\ \hline
		\hline
1	&level\_idc                              	&2	&pic\_width\_in\_mbs\_minus1             	&3	&pic\_height\_in\_map\_units\_minus1     	\\ \hline
4	&frame\_cropping\_flag                   	&5	&frame\_crop\_bottom\_offset             	&6	&log2\_max\_pic\_order\_cnt\_lsb\_minus4 	\\ \hline
7	&num\_ref\_frames                        	&8	&pic\_init\_qp\_minus26                  	&9	&profile\_idc                            	\\ \hline
10	&time\_scale                             	&11	&num\_units\_in\_tick                    	&12	&aspect\_ratio\_info\_present\_flag      	\\ \hline
13	&aspect\_ratio\_idc                      	&14	&log2\_max\_mv\_length\_horizontal       	&15	&log2\_max\_mv\_length\_vertical         	\\ \hline
16	&log2\_max\_frame\_num\_minus4           	&17	&transform\_8x8\_mode\_flag              	&18	&timing\_info\_present\_flag             	\\ \hline
19	&max\_dec\_frame\_buffering              	&20	&entropy\_coding\_mode\_flag             	&21	&constraint\_set0\_flag                  	\\ \hline
22	&pic\_order\_cnt\_type                   	&23	&fixed\_frame\_rate\_flag                	&24	&nal\_hrd\_parameters\_present\_flag     	\\ \hline
25	&direct\_8x8\_inference\_flag            	&26	&vui\_parameters\_present\_flag          	&27	&num\_ref\_idx\_l0\_active\_minus1       	\\ \hline
28	&deblocking\_filter\_control\_present\_flag	&29	&bitstream\_restriction\_flag            	&30	&motion\_vectors\_over\_pic\_boundaries\_flag	\\ \hline
31	&video\_signal\_type\_present\_flag      	&32	&video\_format                           	&33	&colour\_description\_present\_flag      	\\ \hline
34	&colour\_primaries                       	&35	&transfer\_characteristics               	&36	&matrix\_coefficients                    	\\ \hline
37	&pic\_struct\_present\_flag              	&38	&num\_reorder\_frames                    	&39	&max\_bytes\_per\_pic\_denom             	\\ \hline
40	&max\_bits\_per\_mb\_denom               	&41	&chroma\_format\_idc                     	&42	&frame\_mbs\_only\_flag                  	\\ \hline
43	&constraint\_set1\_flag                  	&44	&chroma\_loc\_info\_present\_flag        	&45	&seq\_scaling\_matrix\_present\_flag     	\\ \hline
46	&seq\_scaling\_list\_present\_flag       	&47	&num\_ref\_idx\_l1\_active\_minus1       	&48	&chroma\_qp\_index\_offset               	\\ \hline
49	&frame\_crop\_right\_offset              	&50	&sar\_width                              	&51	&sar\_height                             	\\ \hline
52	&pic\_order\_present\_flag               	&53	&weighted\_pred\_flag                    	&54	&weighted\_bipred\_idc                   	\\ \hline
55	&second\_chroma\_qp\_index\_offset       	&56	&constraint\_set2\_flag                  	&57	&video\_full\_range\_flag                	\\ \hline
58	&overscan\_info\_present\_flag           	&59	&vcl\_hrd\_parameters\_present\_flag     	&60	&mb\_adaptive\_frame\_field\_flag        	\\ \hline
61	&pic\_scaling\_matrix\_present\_flag	    &	&	                                        &   &                                   	    \\ \hline
	\end{tabular}
\end{table*}

\bibliographystyle{IEEEtran}
\bibliography{bibfile}

\end{document}